\documentclass[onecolumn,10pt]{article}
\usepackage{amssymb}
\usepackage{times}
\newcommand{\alg}[1]{\mbox{$\mathfrak{#1}$}}

\newcommand{\hil}[1]{\mbox{$\mathcal{#1}$}}

\newcommand{\ket}[1]{| #1 \rangle}
\newcommand{\bra}[1]{\langle #1 |}

\begin{document}

\title{\textbf{Why the quantum?}}
  \vspace{1in}

\author{\textbf{Jeffrey Bub}\\
\footnotesize Department of Philosophy, University of Maryland,
College Park, MD 20742\footnote{\textit{E-mail address:}
jbub@umd.edu}}

\vspace{.5in}
\date{}

\maketitle

\begin{abstract}
This paper is a commentary on the foundational
significance of the Clifton-Bub-Halvorson theorem characterizing
quantum theory in terms of three information-theoretic constraints. I
argue that: (1) a quantum theory is best understood as a theory
about the possibilities and impossibilities of information transfer, 
as opposed to a theory about the mechanics of nonclassical waves or
particles, (2) given the information-theoretic constraints,
any mechanical theory of quantum phenomena that includes an account of
the measuring instruments that reveal these phenomena must be 
empirically equivalent to a quantum theory, and 
(3) assuming the information-theoretic 
constraints are in fact satisfied in our world, no mechanical theory 
of quantum phenomena that includes an account of measurement 
interactions can be 
acceptable, and the appropriate aim of 
physics at the fundamental level then 
becomes the representation and manipulation of information.
\end{abstract}

\bigskip
\bigskip


\section{Introduction}

This paper is a commentary, as I see it, on the foundational 
significance of the Clifton-Bub-Halvorson (CBH) theorem (Clifton, 
Bub, \& Halvorson, 2003), characterizing quantum theory in terms of 
three information-theoretic constraints. CBH showed that
  one can derive the basic kinematic features of a quantum-theoretic
  description of physical systems---essentially, noncommutativity and
  entanglement---from three fundamental information-theoretic
  constraints: (i) the impossibility of superluminal information transfer
  between two physical systems by performing measurements on one of
  them, (ii) the impossibility of perfectly broadcasting the information
  contained in an unknown physical state (for pure states, this amounts
  to `no cloning'), and (iii) the impossibility of communicating information
so as to implement a certain
  primitive cryptographic protocol, called `bit commitment,'
   with unconditional security. We also partly demonstrated the
   converse derivation, leaving open a question concerning nonlocality
   and  bit commitment. This remaining
  issue has been resolved by Hans Halvorson (Halvorson, 2003a), so we have a
   characterization theorem for quantum theory in terms of the three
   information-theoretic constraints.

I argue for three theses:

  \begin{itemize}
   \item \textit{A quantum theory is best understood as a theory
about the possibilities and impossibilities of information transfer, 
as opposed to a theory about the mechanics of nonclassical waves or
particles.} (By `information' here I mean information in the physical
sense, measured classically by the Shannon entropy or, in a quantum
world, by the von Neumann entropy.)
  \item \textit{Given the information-theoretic constraints,
any mechanical theory of quantum phenomena that includes an account of
the measuring instruments that reveal these phenomena must be 
empirically equivalent to a quantum theory.} 
\item \textit{Assuming the information-theoretic 
constraints are in fact satisfied in our world, no mechanical theory 
of quantum phenomena that includes an account of measurement 
interactions can be 
acceptable, and the appropriate aim of 
physics at the fundamental level 
then becomes the representation and manipulation of information.}
\end{itemize}

 The first thesis follows from the CBH analysis summarized in Section 
 2, and the discussion in 
 Section 3 concerning the problems that arise if one attempts to 
 intepret a quantum theory directly as a nonclassical mechanics. 
 Following CBH, I understand a quantum theory as a theory in which 
 the observables and states have a certain characteristic algebraic 
 structure (just as a relativistic theory is a theory with certain 
 symmetry or invariance properties, defined in terms of a group of 
 space-time transformations). So, for example, the standard quantum 
 mechanics of a system with a finite number of degrees of freedom 
 represented on a single Hilbert space with a unitary dynamics 
 defined by a given Hamiltonian is a 
 quantum theory, and theories with different Hamiltonians can be considered 
 to be empirically inequivalent quantum theories. Quantum field theories 
 for systems with an infinite number 
 of degrees of freedom, where there are many unitarily 
 inequivalent Hilbert space representations of the canonical 
 commutation relations, are quantum theories. 
 
The second thesis is the claim that the information-theoretic
   constraints preclude the possibility of a mechanical theory of
   quantum phenomena, acceptable on empirical grounds, that includes an
   account of the measuring instruments that
   reveal these phenomena. That is, given these constraints, the class of such
   theories is necessarily 
   underdetermined by any empirical evidence. For example, while 
   Bohmian mechanics (Goldstein, 2001)---Bohm's theory with the Born 
   distribution for particle positions---is a perfectly
good candidate for a mechanical theory of quantum phenomena that
includes an account of measurement interactions, 
there can be no empirical grounds for accepting this
version of Bohm's theory
as an answer to the
question that van Fraassen \cite[pp. 2, 242]{Fraassen}
calls `the foundational question \textit{par
excellence}': How could the world possibly be the way a quantum 
theory says it is?)

   The third thesis now follows if we assume that we \textit{do} in fact live
   in a world in which there are certain constraints on the 
   acquisition, representation, and communication of information. I
   argue that the rational epistemological stance in this situation
   is to suspend judgement about the class of empirically equivalent
   but
   necessarily underdetermined mechanical theories that 
   are designed to `solve the
   measurement problem' and regard all these theories as unacceptable.
    In that case, our measuring instruments
   ultimately remain black boxes at some level. This amounts to
   interpreting a quantum theory as a theory about the
   representation and manipulation of information, which then becomes
   the appropriate aim of physics, rather than a theory
   about the ways in which nonclassical waves or particles move.

   The following discussion is divided into three sections:
    `Quantum Theory from Information-Theoretic
  Constraints' (in which I motivate the consideration of these
  particular constraints and, for
   completeness, briefly outline
   the $C^{*}$-algebraic framework in which the CBH characterization theorem
   is formulated), `The Measurement Problem Reconsidered' (in which I 
   review the measurement problem and present 
   arguments for the first two theses), and `The Completeness of Quantum Theory'
   (in which I argue for the third thesis, and show how 
   the information-theoretic characterization of quantum mechanics 
   provides an answer to Wheeler's
   question: `Why the quantum?').

\section{Quantum Theory from Information-Theoretic
  Constraints}

  The question raised by CBH is whether we can deduce the kinematic
  aspects of the quantum-theoretic description of physical systems
  from the assumption that we live in a world in which there are
  certain constraints on the acquisition, representation, and
  communication of information.

  The project was first suggested to me by remarks by Gilles Brassard at
  the meeting `Quantum Foundations in the Light of Quantum Information
  and Cryptography,' held in Montreal, May 17--19, 2000. Brassard and
  Chris Fuchs  (Fuchs, 1997, 2000; Fuchs \& Jacobs, 2002) speculated that
  quantum mechanics could be derived from information-theoretic
constraints formulated in terms of certain primitive cryptographic protocols:
specifically, the possibility of unconditionally secure key 
distribution, and the
impossibility of unconditionally secure bit commitment. I gave a talk
where I mentioned this conjecture,
with some exploration of the motivation for and
background to the `no bit commitment' assumption, at the University of
Pittsburgh Center for Philosophy of Science in December, 2001.
In discussions with Rob Clifton afterwards,
he proposed tackling the problem in the framework of $C^{*}$-algebras,
which eventually led to the CBH paper. A follow-up email message
indicates the excitement we felt at the time:

\begin{quote}
Dec 4, 2001
Jeff---It was good to talk to you over pizza today. In fact, it was the
most exciting `truly
quantum' conversation I've had here with someone since Hans left in July.
  I will definitely try to
organize in my head where we (think!) we are w.r.t. getting the formalism
from no-cloning and
no-commitment (sic)---and I'll summarize it all in an email to you and Hans
in a few days.  In the
meantime, attached is an ecopy of my paper with Hans on entanglement and
open systems (from which I
think you can learn a fair bit about the algebraic formulation of qm) and
the paper you and I were
talking about on noncommutativity and teleportation.
Talk soon.  Rob
\end{quote}

\noindent Although Hans Halvorson was only able to join the project later
(shortly before Clifton's death in August, 2002), he was responsible for a
great deal of the technical
work on the proofs.

A $C^{*}$-algebra (as I learned!) is essentially an abstract 
generalization of the
structure of the algebra of operators on a Hilbert space.
Technically, a (unital) $C^{*}$-algebra is a Banach $^{*}$-algebra over the
complex numbers containing the identity, where the involution
operation $^{*}$ and the norm are related by $\|A^{*}A\| = \|A\|^{2}$.
So the algebra $\alg{B}(\hil{H})$ of all bounded operators on a
Hilbert space
$\hil{H}$ is a $C^{*}$-algebra,
with $^{*}$
the adjoint operation and $\|\cdot\|$ the
standard operator norm.

In standard quantum theory, a state on $\alg{B}(\hil{H})$ is
defined by
a density operator $D$ on $\hil{H}$ in terms of an expectation-valued
functional
$\rho(A)=\mbox{Tr}(AD)$ for all observables represented by
self-adjoint operators $A$ in $\alg{B}(\hil{H})$. This definition of $\rho(A)$
in terms of $D$
yields a positive normalized linear functional. So a
  state on a $C^{*}$-algebra $\alg{C}$ is defined, quite generally, as
any positive normalized
linear functional $\rho:\alg{C}\rightarrow\mathbb{C}$ on the algebra.
Pure states are defined by the condition that if
$\rho=\lambda\rho_{1}+(1-\lambda)\rho_{2}$ with $\lambda\in (0,1)$,
then
$\rho=\rho_{1}=\rho_{2}$;  other states are mixed.
A pure state
in standard quantum theory corresponds to a density operator for which
$D^{2}=D$, and this is equivalent to the existence of a unit vector
$\ket{v}\in\hil{H}$ representing the state of the system via
$\rho(A)=\bra{v}A\ket{v}$.  In a  $C^{*}$-algebra, since countable
additivity is not presupposed by the $C^{*}$-algebraic notion of 
state (and, theorefore, Gleason's theorem does not apply),
   there can be pure states of $\alg{B}(\hil{H})$ in the 
   $C^{*}$-algebraic sense that are not
  representable by vectors in $\hil{H}$ (nor by density operators in $\hil{H}$).

The most general dynamical evolution of a system
represented by a $C^{*}$-algebra of observables is given by a
completely positive linear map $T$ on the algebra of observables,
where $0\leq T(I)\leq I$. The map or operation $T$ is called selective
if $T(I)<I$
and nonselective if $T(I)=I$. A yes-no measurement of some idempotent
observable represented by a projection operator $P$ is an example of a
selective operation. Here,  $T(A)=PAP$
for all $A$ in the $C^{*}$-algebra $\alg{C}$,
and $\rho^{T}$, the transformed (`collapsed') state, is the final state
obtained after measuring $P$ in the
state $\rho$
  and ignoring all elements of the ensemble that do not
yield the eigenvalue 1 of $P$
(so $\rho^{T}(A)=\rho(T(A))/\rho(T(I))$ when $\rho(T(I))\not=0$, and
$\rho^{T}=0$ otherwise). The time evolution in the Heisenberg picture
induced by a unitary operator $U\in\alg{C}$ is an example of a
nonselective operation. Here, $T(A)=UAU^{-1}$. Similarly, the
measurement of an observable $O$ with spectral measure $\{P_{i}\}$,
without selecting a particular outcome, is an example of a
nonselective operation, with $T(A) = \sum _{i=1}^{n}P_{i}AP_{i}$. Note 
that any completely positive linear map can be regarded as the 
restriction to a local 
system of a unitary map on a larger system.

A
representation of a $C^{*}$-algebra $\alg{C}$ is any mapping
$\pi:\alg{C}\rightarrow\alg{B}(\hil{H})$ that preserves the linear,
product, and $^{*}$ structure of \alg{C}. The representation is faithful
if $\pi$ is
one-to-one, in which case $\pi(\alg{C})$ is an isomorphic copy of
$\alg{C}$. The Gelfand-Naimark theorem says that every abstract $C^{*}$-algebra
has a concrete
faithful representation as a norm-closed $^{*}$-subalgebra of
$\alg{B}(\hil{H})$, for some appropriate Hilbert space $\hil{H}$.
In the case of systems with an
infinite number of degrees of freedom  (as in quantum field theory),
it turns out that there are inequivalent representations of the
$C^{*}$-algebra of observables defined by the commutation relations.

Apart from this infinite case, it might seem that $C^{*}$-algebras offer no
more than an abstract way of talking about quantum mechanics. In
fact, the $C^{*}$-algebraic formalism provides a mathematically
abstract characterization of a broad class of physical theories that
includes all classical mechanical particle and field theories, as well as
quantum mechanical theories. One could, of course, consider weaker
mathematical structures (such as Jordan-Banach algebras, or Segal 
algebras (Segal, 1947)), but it seems that that the $C^{*}$-algebraic
machinery suffices for all physical theories that have been found to
be empirically successful to date, including phase space theories and
Hilbert space theories (Landsman, 1998), and theories based on a
manifold (Connes, 1994).

The relation between classical theories and $C^{*}$-algebras is
this: every
\textit{commutative} $C^{*}$-algebra $\alg{C}$ is isomorphic to the set
$C(X)$ of
all continuous complex-valued functions on a locally
compact Hausdorff space
$X$ that go to zero at infinity. 
If $\alg{C}$ has a multiplicative identity, $X$ is compact.
So behind every abstract commutative $C^{*}$-algebra there is a
classical phase space theory defined by this `function representation'
on the phase space $X$. Conversely, every classical phase space theory
defines a $C^{*}$-algebra. For example, the observables of a classical
system of $n$ point particles---real-valued functions on the phase space
$\mathbb{R}^{6n}$---can be represented as the self-adjoint elements of
the $C^{*}$-algebra $\alg{B}(\mathbb{R}^{6n})$ of all continuous complex-valued
functions $f$ on $\mathbb{R}^{6n}$ that go to zero at infinity. The phase space
$\mathbb{R}^{6n}$ is only locally compact (so 
$\alg{B}(\mathbb{R}^{6n})$ does not have a multiplicative identity), 
but it can be made compact by
adding just one point `at infinity,' or we can simply consider a
bounded (and thus compact) subset of $\mathbb{R}^{6n}$.
The statistical states of
the system are given by probability
measures $\mu$ on
$\mathbb{R}^{6n}$, and pure states, corresponding to maximally complete
information about the particles, are given by
the individual points of
$\mathbb{R}^{6n}$. The system's state $\rho$ in
the $C^{*}$-algebraic sense is the
expectation functional corresponding to $\mu$, defined by
$\rho(f)=\int_{\mathbb{R}^{6n}}f\mbox{d}\mu$.

So classical theories are characterized  by commutative
$C^{*}$-algebras. The question is whether quantum theories should be
identified with the class of noncommutative $C^{*}$-algebras, or with
some appropriate subclass.

Before tackling this question, it will be worthwhile to clarify the
significance of the two
information-theoretic principles: `no superluminal information transfer
via measurement,' and `no broadcasting.'

Consider a
composite quantum system A+B, consisting of two subsystems, A and B.
For simplicity, assume the systems are identical, so their
$C^{*}$-algebras $\alg{A}$ and $\alg{B}$ are isomorphic.
The
observables of the component systems A and B are
represented by the self-adjoint elements of $\alg{A}$ and $\alg{B}$,
respectively. Let  $\alg{A}\vee\alg{B}$ denote the $C^{*}$-algebra
generated by $\alg{A}$ and $\alg{B}$. The physical states of A, B, and
A+B, are given by positive normalized linear functionals on their
respective algebras that encode the expectation values of all
observables. To capture the idea that A and B are \textit{physically
distinct} systems,
we assume (as a necessary condition) that any state of $\alg{A}$ is
compatible with any state of $\alg{B}$, i.e., for any state
$\rho _{A}$ of $\alg{A}$ and $\rho _{B}$ of $\alg{B}$,
there is a state $\rho$ of $\alg{A}\vee \alg{B}$ such that $\rho
|_{\alg{A}}=\rho _{A}$ and $\rho |_{\alg{B}}=\rho _{B}$.

The sense of the `no superluminal information transfer via 
measurement' constraint
is
that when Alice and Bob, say, perform local measurements, Alice's
measurements can have no influence on the statistics for the outcomes
of Bob's measurements, and conversely. That is, merely performing
  a local measurement---in the nonselective sense---cannot, in and of
itself, convey any information to a physically distinct system, so
that everything `looks the same' to that system after the
measurement operation as before, in terms of the expectation values
for its own local observables. (The restriction to
nonselective measurements is required here, of course, because
selective measurement operations will in general change the statistics
of observables measured at a distance, simply because the ensemble
relative to which the statistics is computed changes with the
selection.) It follows from this constraint that A and B are
\textit{kinematically independent} systems if they are physically
distinct in the above sense, i.e., every element of $\alg{A}$
commutes pairwise with every element of $\alg{B}$.

The `no broadcasting' condition now ensures that the individual algebras
$\alg{A}$ and  $\alg{B}$ are noncommutative. Broadcasting is a
process closely related to cloning. In fact, for pure states, broadcasting
reduces to cloning.  In cloning, a ready state $\sigma$ of
a system B and the state to be cloned $\rho$ of system A are
transformed into two copies of $\rho$. In broadcasting, a ready state
$\sigma$ of B and the state to be broadcast $\rho$ of A are 
transformed to a new
state $\omega$ of A+B, where the marginal states of $\omega$ with
respect to both A and B are $\rho$. In elementary quantum
mechanics, neither cloning nor broadcasting is possible in general. A
pair of pure states can be cloned if and only if they are orthogonal
and, more generally, a pair of mixed states can be broadcast if
and only if they are represented by mutually commuting density
operators. In CBH, we show that broadcasting and cloning are always 
possible for
classical systems, i.e., in the commutative case there is a
universal broadcasting map that clones
any pair of input pure states and broadcasts any pair of input mixed
states. Conversely, we show that if any two states can be (perfectly) 
broadcast,
then any two
pure states can be cloned; and if two pure states of a $C^{*}$-algebra
can be cloned, then they must be orthogonal. So, if any
two states can be broadcast, then all pure states are orthogonal,
from which it follows that the algebra is commutative.

So far, we have the following: For a composite system A+B, the
`no superluminal information transfer via measurement'
constraint entails that the
$C^{*}$-algebras $\alg{A}$ and $\alg{B}$, whose self-adjoint elements
represent the observables of A and B,
commute with each other; and the `no broadcasting'
constraint
entails that the algebras $\alg{A}$ and $\alg{B}$ separately are 
noncommutative. The quantum mechanical phenomenon of interference is the
physical manifestation of the noncommutativity of quantum 
observables or, equivalently, the superposition
of quantum states. From the above analysis, we
see that the impossibility of perfectly broadcasting the
information contained in an unknown physical state, or of cloning or
copying the information in an unknown pure state, is the
information-theoretic counterpart of interference.

To return to the question at issue: 
if $\alg{A}$ and $\alg{B}$ are noncommutative and mutually commuting,
it can be shown that there are nonlocal entangled states on the
$C^{*}$-algebra
$\alg{A}\vee\alg{B}$ they generate (see Landau, 1987;
Summers, 1990;
Bacciagaluppi, 1993; and---more relevantly here, in
terms of a specification of the range of entangled states that can be
guaranteed to exist---Halvorson, 2003a). So it seems that
entanglement---what Schr\"{o}dinger (1935, p. 555) called
  `\textit{the} characteristic trait of
quantum mechanics, the one that enforces its entire departure from
classical lines of thought'--- follows automatically in
any theory with a noncommutative algebra of observables. That is, it
seems that once we assume `no superluminal information transfer via
measurement,' and `no broadcasting,' the class of allowable physical
theories is restricted to those theories in which physical systems manifest
both interference \textit{and} nonlocal entanglement. So if we take 
interference and nonlocal entanglement as the characteristic physical 
attributes that distinguish quantum systems from classical systems, it 
might seem that we should simply identify quantum theories with the 
class of noncommutative $C^{*}$-algebras. 

This
conclusion is surely too quick, though, since the derivation of entangled 
states depends on formal
properties of the
$C^{*}$-algebraic machinery. Suppose we considered more general 
algebraic structures, such as Segal algebras (see Segal, 1947), 
which have the minimal amount of 
structure required for spectral theory (i.e., the minimal structure 
needed to make sense of the probabilities of measurement outcomes). 
Hans Halvorson (Halvorson, 2003c) has speculated that the 
existence of entangled states would not follow from `no superluminal 
information transfer' and `no broadcasting' in Segal algebras, but 
would require, in addition, the `no bit commitment' constraint. 
(To show this is a future project.)  In an
information-theoretic characterization of quantum theory, the fact that
entangled states can be instantiated, and instantiated nonlocally,
should be shown to follow from  some
information-theoretic principle. The role of the `no bit
commitment' constraint is to guarantee that nothing prevents a certain range of
nonlocal entangled states from being instantiated in our
world---that physical systems can be prepared in such states.

To motivate this principle, consider Schr\"{o}dinger's discussion of
entanglement in his extended two-part
commentary (Schr\"{o}dinger, 1935, 1936)
on the
Einstein-Podolsky-Rosen (EPR) argument (Einstein, Podolsky, \& Rosen,
1935).

In the
first part, Schr\"{o}dinger considers entangled states for which the 
biorthogonal
decomposition is unique, as well as cases like the EPR-state, where
the
biorthogonal decomposition is non-unique. There he is concerned to
show that suitable measurements on one system can fix the (pure)
state of the
entangled distant system, and that this state depends on what
observable one chooses to measure, not merely on the outcome of that
measurement. In the second part, he shows that a
`sophisticated experimenter,' by performing a suitable local
measurement on
one system, can `steer' the
distant system into any mixture of pure states representable by
its reduced density
operator. (So the distant
system can be steered into any pure state in the support of the
reduced density operator, with a nonzero probability that depends
only
on the pure state.) For a mixture of linearly
independent states, the steering can be done by performing a
PV-measurement in a
suitable basis. If the states are linearly dependent, the
experimenter
performs what we would now call a POV-measurement, which amounts to enlarging the
experimenter's Hilbert space by adding an ancilla, so that the
dimension of the enlarged Hilbert space is equal to the number of
linearly dependent states.

For example, suppose Alice and Bob each hold one of a pair of
spin-$\frac{1}{2}$
particles in the entangled EPR state:
  \[ \ket{\psi} = \frac{1}{\sqrt{2}}(\ket{+}_{A}\ket{-}_{B}-
  \ket{-}_{A}\ket{+}_{B}) \]
  where $\ket{+}$ and $\ket{-}$ are the
eigenstates of the Pauli spin operator $\sigma_{z}$.

Bob's state is represented by the density operator $\rho_{B} =
\frac{1}{2}I$.
This can be interpreted as an equal weight mixture of the states
$\ket{+}_{B}$, $\ket{-}_{B}$,
   but also as an infinity of other
mixtures including, to take a specific example,
the equal weight mixture of the four
nonorthogonal states:
\begin{eqnarray}
\ket{\phi_{1}}_{B} &=& \alpha\ket{+}_{B} + \beta\ket{-}_{B} \nonumber \\
\ket{\phi_{2}}_{B} &=& \alpha\ket{+}_{B} - \beta\ket{-}_{B} \nonumber \\
\ket{\phi_{3}}_{B} &=& \beta\ket{+}_{B} + \alpha\ket{-}_{B} \nonumber \\
\ket{\phi_{4}}_{B} &=& \beta\ket{+}_{B} - \alpha\ket{-}_{B} \nonumber
\end{eqnarray}
That is:
\[\rho_{B} = \frac{1}{4}(\ket{\phi_{1}}\bra{\phi_{1}} +
\ket{\phi_{2}}\bra{\phi_{2}} + \ket{\phi_{3}}\bra{\phi_{3}} +
\ket{\phi_{4}}\bra{\phi_{4}}) \nonumber \\
= \frac{1}{2}I \nonumber \]

If Alice measures the spin observable with eigenstates $\ket{+}_{A}$,
$\ket{-}_{A}$ on her particle A and Bob measures the corresponding
spin observable on his particle B, Alice's outcomes will be oppositely
correlated with Bob's outcomes (+ with -, and - with +). If,
instead, Alice prepares a spin-$\frac{1}{2}$ ancilla particle
A$^{\prime}$ in the state
$\ket{\phi_{1}}_{A^{\prime}} = 
\alpha\ket{+}_{A^{\prime}} + \beta\ket{-}_{A^{\prime}}$ 
and measures an observable on the pair of
systems A+A$^{\prime}$ in her possession with eigenstates:
\begin{eqnarray}
\ket{1} &=& \frac{1}{\sqrt{2}}(\ket{+}_{A^{\prime}}\ket{-}_{A} -
  \ket{-}_{A^{\prime}}\ket{+}_{A}) \nonumber \\
  \ket{2} &=& \frac{1}{\sqrt{2}}(\ket{+}_{A^{\prime}}\ket{-}_{A} +
  \ket{-}_{A^{\prime}}\ket{+}_{A}) \nonumber \\
  \ket{3} &=& \frac{1}{\sqrt{2}}(\ket{+}_{A^{\prime}}\ket{+}_{A} -
  \ket{-}_{A^{\prime}}\ket{-}_{A}) \nonumber \\
  \ket{4} &=& \frac{1}{\sqrt{2}}(\ket{+}_{A^{\prime}}\ket{+}_{A} +
  \ket{-}_{A^{\prime}}\ket{-}_{A}) \nonumber
  \end{eqnarray}
(the Bell states), she will obtain the outcomes 1, 2, 3, 4
with equal probability, and these outcomes will be correlated with Bob's
states $\ket{\phi_{1}}_{B}$, $\ket{\phi_{2}}_{B}$,
$\ket{\phi_{3}}_{B}$, $\ket{\phi_{4}}_{B}$ (i.e., if Bob
checks to see whether his particle is in the state
$\ket{\phi_{i}}_{B}$ when Alice reports that she obtained the outcome
$i$, he will find that this is always in fact the case). This follows
because:
\[
\ket{\phi_{1}}_{A^{\prime}}\ket{\psi} =
\frac{1}{2}(-\ket{1}\ket{\phi_{1}}_{B} -
\ket{2}\ket{\phi_{2}}_{B} + \ket{3}\ket{\phi_{3}}_{B} +
\ket{4}\ket{\phi_{4}}_{B}) \nonumber \]
\noindent In this sense, Alice can steer Bob's particle into any mixture
compatible with the density operator $\rho_{B} = \frac{1}{2}I$ by an
appropriate local measurement.

What Schr\"{o}dinger found problematic about
entanglement was the possibility of remote steering
(Schr\"{o}dinger, 1935, p. 556):
\begin{quote}
It is rather discomforting that the theory should allow a system to
be steered or piloted into one or the other type of state at the
experimenter's mercy in spite of his having no access to it.
\end{quote}

Notice that remote steering in this probabilistic sense is precisely
what makes quantum teleportation possible. Suppose Alice and Bob share
a pair of spin-$\frac{1}{2}$ particles A and B in
the EPR state and Alice is given a
spin-$\frac{1}{2}$ particle A$^{\prime}$
  in an \textit{unknown} state $\ket{\phi_{1}}$. If
Alice measures the composite system A+A$^{\prime}$
in the Bell basis, she will steer Bob's particle into
one of the states $\ket{\phi_{1}}_{B}$, $\ket{\phi_{2}}_{B}$,
$\ket{\phi_{3}}_{B}$, $\ket{\phi_{4}}_{B}$ with equal probability. If
Alice tells Bob the outcome of her measurement, Bob can apply a local
unitary transformation to obtain the state $\ket{\phi_{1}}_{B}$:
\begin{itemize}
\item[] 1: apply the transformation $I$ (the identity---i.e., do
nothing)
\item[] 2: apply the transformation $\sigma_{z}$
\item[] 3: apply the transformation $\sigma_{x}$
\item[] 4: apply the transformation $-i\sigma_{y}$
\end{itemize}

Today we know that remote steering and nonlocal entanglement are physically
possible, but in 1936 Schr\"{o}dinger conjectured that an entangled state
of a composite system might
decay to a mixture as soon as the component systems
separated. So while
there would still be correlations between the states of the
component systems, remote steering would no longer be possible
(Schr\"{o}dinger, 1936, p. 451):
\begin{quote}
     It seems worth noticing that the [EPR] paradox could be avoided
by a
     very simple assumption, namely if the situation after separating
     were described by the expansion (12), but with the additional
     statement that the knowledge of the \textit{phase relations}
     between the complex constants $a_{k}$ has been entirely lost in
     consequence of the process of separation. This would mean that
     not only the parts, but the whole system, would be in the
     situation of a mixture, not of a pure state. It would not preclude
     the possibility of determining the state of the first system by
     \textit{suitable} measurements in the second one or \textit{vice
     versa}. But it would utterly eliminate the experimenters influence
     on the state of that system which he does not touch.
\end{quote}

\noindent Expansion (12) is the biorthogonal expansion:
\noindent
\begin{equation}
\Psi(x,y) = \sum_{k}a_{k}g_{k}(x)f_{k}(y) \nonumber
\end{equation}

It seems that Schr\"{o}dinger regarded the phenomenon of
interference associated with noncommutativity
in quantum mechanics as unproblematic, because he saw this as
reflecting the fact that particles are
wavelike.
But he did not believe that we live
in a world in which physical systems can exist nonlocally in
entangled
states, because such states would allow remote steering, i.e.,
effectively teleportation.
Schr\"{o}dinger did not expect that
experiments would bear this out and thought that
nonlocal entangled states were simply an artifact of the formalism
(like paraparticle states, which are allowed in Hilbert space quantum
mechanics but not observed in nature).

Schr\"{o}dinger's conjecture
raises the possibility of a quantum-like world in which there is
interference but
no nonlocal entanglement, and this possibility needs to be excluded on
information-theoretic grounds. This is the function of the `no bit
commitment' constraint.

Bit commitment is a cryptographic protocol in which one party, Alice,
supplies an encoded bit
to a second party, Bob, as a warrant for her commitment to 0 or 1.
The information available in the encoding
should be insufficient for Bob to ascertain the value of the bit at
the initial commitment stage, but
sufficient, together with further information supplied by Alice at a
later stage when she is supposed to `open' the commitment by revealing the
value of the bit, for Bob to be convinced that the protocol does not
allow Alice to cheat by encoding the bit in a way that leaves her
free
to reveal either 0 or 1 at will.

In 1984, Bennett and Brassard (Bennett \& Brassard, 1984) proposed a 
quantum bit
commitment protocol now referred to as BB84. The basic idea was to
encode
the 0 and 1
commitments as two quantum mechanical mixtures
represented by the same density operator, $\omega$. As they showed,
Alice can cheat by adopting an
EPR
attack or cheating strategy. Instead of following the protocol and
sending a particular mixture to Bob she prepares pairs of
particles A+B in the same entangled state $\rho$, where
$\rho |_{\alg{B}} = \omega$. She keeps one of each pair (the ancilla A)
and sends the second
particle B to Bob, so that Bob's particles are in the mixed state
$\omega$. In this way she can reveal either bit at will at the 
opening stage, by
effectively steering Bob's particles into
the desired mixture via appropriate measurements on her ancillas. Bob
cannot detect this cheating strategy.

Mayers (1996, 1997), and Lo and Chau (1997), showed that
the insight of
Bennett and Brassard
can be extended to a proof that a generalized version of
the EPR cheating strategy can always be applied, if the
Hilbert space is enlarged in a suitable way by introducing additional
ancilla particles. The proof of this `no go' quantum bit commitment
theorem
exploits biorthogonal decomposition
via a result by Hughston, Jozsa, and Wootters (1993)
(effectively anticipated by
Schr\"{o}dinger's analysis). Informally, this  says that for a quantum
mechanical system
consisting of two (separated) subsystems represented by
  the $C^{*}$-algebra
$\alg{B}(\hil{H}_{1}) \otimes \alg{B}(\hil{H}_{2})$, any mixture of
states on $\alg{B}(\hil{H}_{2})$ can be generated
from a distance by performing an appropriate POV-measurement on the
system represented by $\alg{B}(\hil{H}_{1})$, for an appropriate
entangled state of the composite system
  $\alg{B}(\hil{H}_{1}) \otimes \alg{B}(\hil{H}_{2})$.
This is what makes it possible for Alice to cheat in her bit
commitment protocol with Bob. It is easy enough to see this for the
original BB84 protocol. Suprisingly, this is also the case for
  any conceivable quantum bit commitment
protocol. See Bub (2001) for a discussion.

Now, unconditionally secure bit commitment is also impossible for
classical
systems, in which the algebras of observables are
commutative.\footnote{Adrian Kent (1999) has shown
how to implement a secure classical
bit commitment protocol by exploiting relativistic signalling
constraints in a timed sequence of communications between verifiably
separated sites for both Alice and Bob. In a bit commitment protocol,
as usually construed, there is a time interval
of arbitrary
length, where no information is exchanged,
between the end of the commitment stage of the protocol and
the opening or unveiling stage, when Alice reveals the value of the
bit.
  Kent's ingenious scheme effectively involves a third stage between the
commitment state and the unveiling stage, in which information is
exchanged between Bob's sites and Alice's sites at regular intervals
until one of Alice's sites
chooses to unveil the originally committed bit. At this moment of
unveiling the protocol is not yet complete, because a further sequence of
unveilings is required between Alice's sites and corresponding sites
of Bob before Bob has all the information required to verify the
commitment at a single site. If a bit commitment protocol
is understood to
require an arbitrary amount of `free' time between the end of the
commitment stage and the opening stage (in which no step is to be
executed in the protocol), then unconditionally secure bit commitment
is impossible for classical systems.
(I am indebted to Dominic Mayers for clarifying this
point.)} But the insecurity of any bit commitment protocol
in a noncommutative setting depends on considerations entirely
different from those in a classical commutative setting. Classically,
unconditionally secure bit commitment is impossible, essentially
because
  Alice can send (encrypted) information to Bob that guarantees the
  truth of an exclusive classical disjunction (equivalent
to her commitment to a 0 or a 1) only if the information is biased
towards one
of the alternative disjuncts (because a classical exclusive
disjunction is true
if and only if one of the disjuncts is true and the other false). No
principle of classical mechanics precludes Bob from extracting this
information.
So the security of the protocol cannot be unconditional and
  can only depend on issues of
computational complexity.

  By contrast, in a
situation of the sort envisaged by Schr\"{o}dinger, in which the
algebras of observables are noncommutative but composite physical systems
cannot exist in nonlocal entangled states,
  if Alice sends Bob one of two mixtures associated with the same
  density operator to establish her commitment, then she
is, in effect, sending Bob evidence for the
truth of an exclusive disjunction that is not based on the selection
of a particular disjunct. (Bob's reduced density operator
is associated ambiguously with both mixtures, and
hence with the truth of the exclusive disjunction: `0 or 1'.)
Noncommutativity allows the possibility of different mixtures associated with
the same density operator. What thwarts the possibility of using the
ambiguity of mixtures in this way to implement an unconditionally
secure bit commitment protocol is the existence of nonlocal entangled
states between Alice and Bob. This allows Alice to cheat by preparing
a suitable entangled state instead of one of the mixtures, where the
reduced density operator for Bob is the same as that of the mixture.
Alice is
then able to steer Bob's systems into either of the two mixtures
associated with the alternative commitments at will.

So what \textit{would} allow unconditionally secure bit commitment in a
noncommutative theory is the absence of physically occupied
  nonlocal entangled states.
One can therefore take Schr\"{o}dinger's remarks as relevant to the
question of whether or not secure bit
commitment is possible in our world. In effect, Schr\"{o}dinger
raises the possibility
that we live in a quantum-like world in which  secure bit commitment
is possible! The suggestion is that if Alice and Bob prepare
two particles A+B in an
entangled state whose
biorthogonal decomposition is:
\[\ket{\psi} = \sum \sqrt{\lambda_{i}}\ket{a_{i}}\ket{b_{i}}\]
and then separate, each taking one particle,
the phase relations between the components of the
density operator of the composite system $\rho =
\ket{\psi}\bra{\psi}$ will become randomized (presumably, virtually
instantaneously), resulting in the transition:
\[\rho \longrightarrow
\sum \lambda_{i}\ket{a_{i}}\bra{a_{i}}\otimes\ket{b_{i}}\bra{b_{i}}\]
so that:
\begin{eqnarray}
\rho_{A} &=& \sum\lambda_{i}\ket{a_{i}}\bra{a_{i}} \\
\rho_{B} &=& \sum\lambda_{i}\ket{b_{i}}\bra{b_{i}}
\end{eqnarray}
Then unconditionally secure bit commitment would be possible. Alice would
have to prepare a specific mixture associated with a particular
commitment---she could
no longer steer Bob's particles at will into one of two alternative
mixtures consistent with the same density operator by exploiting the
EPR cheating strategy. It follows that the
impossibility of unconditionally secure bit commitment entails that,
for any mixed state that Alice and Bob can prepare by following some
(bit commitment) protocol, there is a corresponding nonlocal
entangled state that can be physically occupied by Alice's and Bob's
particles.

What CBH showed was that quantum theories---theories where 
 (i) the observables of the theory are represented by the self-adjoint 
 operators in a 
 noncommutative $C^{*}$-algebra (but the algebras of observables of 
 distinct systems commute), (ii) the states of the theory are represented 
 by $C^{*}$-algebraic states (positive normalized linear functionals 
 on the $C^{*}$-algebra), and spacelike separated systems can be 
 prepared in
entangled states that allow remote steering, and (iii) dynamical 
changes are 
represented by 
 completely positive linear maps---are characterized by the
three information-theoretic `no-go's': no superluminal communication
of information via measurement, no (perfect) broadcasting, and no
(unconditionally secure) bit commitment.

\section{The Measurement Problem Reconsidered}

A $C^{*}$-algebra is, in the first instance, relevant to physical theory
as an algebra of observables, and states defined as expectation-valued
functionals over these observables. Observables here are to be
constrasted with `beables' in Bell's terminology (Bell, 1987a) or dynamical
quantities, where the idempotent dynamical quantities correspond to
properties of physical systems, and the $C^{*}$-algebraic states assign
probabilities to ranges of values of observables and (unlike
classical states) do not represent complete catalogues of properties.

The picture, broadly speaking, is this: At the start of a physical
investigation, one begins by making measurements with instruments
that have the status of black boxes relative to
the future theory that will eventually arise out of the
investigation. Of course, the instruments (and their inputs and
outputs) will be described in terms
of current theory, whatever that is, but at this stage (since we are
supposing that the current theory will be replaced) the instruments are,
epistemologically, just black
boxes that we use to investigate statistical correlations.
David Albert's book (Albert, 1992),
\textit{Quantum Mechanics and Experience}, begins the account of
quantum phenomena in this way, with instruments called `colour' boxes
and `hardness' boxes that are essentially black boxes of
different types that take
a system in an input state (the output of another black box)
  and produce a system in one of two
output states, with a certain probability that depends on the input 
state (they correspond to instruments for measuring the spin
of an electron in different directions). One investigates the
statistics produced by these black boxes in various combinations and
arrives (creatively, not inductively) at a certain algebraic
structure for the observables and probabilistic states associated
with the systems, and a dynamics that accounts for change
between measurements. To say that the algebraic structure is a
$C^{*}$-algebra is just to impose certain minimal formal constraints on
the structure of observables and states that, we expect, will be
applicable to any
physical theory that we might want to consider (and these constraints 
do in fact
characterize all physical theories that have been considered in the
past 400 years or so). For example, the $C^{*}$-algebraic constraints
exclude haecceitist theories that associate a primitive `thisness'
with physical systems. (See the discussion by Halvorson (2003b) and
by Halvorson and Bub (2003) on toy theories proposed by Smolin (2003) and by
Spekkens (2003) that are not $C^{*}$-algebraic
theories.)
We might, of course, at some point have good reasons to
consider a broader class of algebraic structures than 
$C^{*}$-algebras (e.g., Segal algebras), 
and the discussion here is not intended to exlude this
possibility.\footnote{By imposing the three information-theoretic
constraints on $C^{*}$-algebras, we characterize a class of quantum
theories with representations in complex Hilbert spaces. One would
like to rule out real and quaternionic Hilbert spaces on
information-theoretic grounds as well, so this in itself would
suggest broadening the class of algebraic structures.}
For the three theses about quantum theory argued for here,
it is sufficient to note that
$C^{*}$-algebras characterize a broad class of theories including all
present and past classical and quantum theories of both field and 
particle varieties,
and hybrids of these theories (for example, theories with
superselection rules).

So suppose we arrive at a theory formulated in this way in terms of
a $C^{*}$-algebra of observables and states. There are two cases to
consider. If the algebra is commutative, there is a phase space
representation of the theory---not necessarily the phase space of
classical mechanics, but a theory in which the observables of the
$C^{*}$-algebra are replaced by `beables' or dynamical quantities,
and the $C^{*}$-algebraic states are replaced by states representing complete
catalogues of properties (idempotent quantities). In this case, it is
possible to extend the theory to include the measuring instruments
that are the source of the $C^{*}$-algebraic statistics, so that they are no
longer black boxes but constructed out of systems that are
characterized by properties and states of the phase space theory.
That is, the $C^{*}$-algebraic theory can be replaced by a `detached observer'
theory of the
physical processes underlying the phenomena, to use Pauli's term
(Pauli, 1954),
including the processes involved in the functioning of measuring
instruments.

Note that this depends on a representation theorem. In the
noncommutative case, we are guaranteed only the existence of a
Hilbert space representation of the $C^{*}$-algebra, and it is an open
question whether a `detached observer' description of the phenomena is
possible.

In the case of a quantum theory, suppose we interpret the
Hilbert space representation as the
noncommutative analogue of a phase space theory. That is, suppose we
interpret the quantum
state of a system as providing a complete catalogue of the system's
properties---as complete as possible in a noncommutative
setting (so the catalogue includes all the properties represented
by projection operators assigned
unit probability by the state). As Einstein realized, such an
interpretation runs into trouble because of the existence of entangled
states. In a 1948 letter to
Max Born, he writes (Born, 1971, p. 164):
\begin{quote}
I just want to explain what I mean when I say that we should try to
hold on to physical reality. We all of us have some idea of what the
basic axioms in physics will turn out to be. The quantum or the
particle will surely not be amongst them; the field, in Faraday's or
Maxwell's sense, could possibly be, but it is not certain. But
whatever we regard as existing (real) should somehow be localized in
time and space. That is, the real in part of space $A$ should (in
theory) somehow `exist' independently of what is thought of as real
in space $B$. When a system in physics extends over the parts of
space $A$ \textit{and} $B$, then that which exists in $B$ should
somehow exist independently of that which exists in $A$. That which really
exists in $B$ should therefore not depend on what kind of measurement
is carried out in part of space $A$; it should also be independent of
whether or not any measurement at all is carried out in space $A$. If
one adheres to this programme, one can hardly consider the
quantum-theoretical description as a complete representation of the
physically real. If one tries to do so in spite of this, one has to
assume that the physically real in $B$ suffers a sudden change as a
result of a measurement in $A$. My instinct for physics brisles at
this. However, if one abandons the assumption that what exists in
different parts of space has its own, independent, real existence,
then I simply cannot see what it is that physics is meant to describe.
For what is thought to be a `system' is, after all, just a convention,
and I cannot see how one could divide the world objectively in such a
way that one could make statements about parts of it.
\end{quote}

The problem, for Einstein, is a conflict with two principles that he
regarded as crucial for realism: separability (the world can be
divided into separable systems with their own properties: what we
think of as existing or real in region $A$ should exist independently of
what we think of as existing or real in region $B$), and locality
(the properties of a system in region $A$ should be independent of what we
choose to measure in region $B$, or whether any measurement at all is
performed in region $B$). Now, the possibility of entangled
states over any pair of spatially separated regions $A$ and $B$ means
that a measurement at $A$ can
change the catalogue of properties not only at $A$ but also at $B$,
and this violates locality. Alternatively, if we assume that a system in
region B does not have any properties independently of the properties
of system A, then we violate separability. The separability and
locality conditions, formulated as constraints on probabilities,
are equivalent to the
assumption that correlations can be reduced to a common cause, and
Bell's derivation of an inequality (violated by certain quantum
correlations) from these conditions is an
elegant demonstration of a surprising implication of Einstein's insight: the
impossibility of embedding the quantum correlations in a common cause
theory.

Aside from this difficulty, there is a further problem
associated with entangled states in carrying
through this interpretation of the Hilbert space theory as a `detached
observer' theory. If we take the
quantum state of a system as providing a complete catalogue of the
properties of the system  (all the properties represented
by projection operators assigned
unit probability by the state), then a unitary dynamics 
(which is linear in the sense 
that superpositions of vector states are mapped onto corresponding 
superpositions of image vector states) entails that a measuring instrument
will generally end up entangled with the system it
measures. So at the end of what we take to be a
measurement,
  neither the measuring instrument nor the system measured
will have separable properties associated with our commonsense account of
the phenomenon (that the instrument registers a definite outcome,
associated with a definite property of the system).  This is the
measurement problem, or the problem of Schr\"{o}dinger's cat (where
the cat plays the role of a macroscopic measuring instrument):
it is impossible to extend the Hilbert space theory as a
noncommutative mechanics to include the black box measuring
instruments.

The orthodox response to this problem is the proposal that the
unitary dynamics is suspended whenever a quantum system is measured,
and that the problematic entangled state `collapses' to one of the terms in the
superposition, the term corresponding to the registration of a
definite outcome (so the final quantum state at the end of a
measurement is represented as a mixed state over the different
outcomes, with weights equal to the
probabilities defined by the entangled state). But this response is
inadequate without an account, in physical terms, of
what distinguishes measurements
from other physical processes. Without such an account, measuring
instruments are still black boxes and we do not have a `detached
observer' theory.

`Collapse' theories like the GRW theory (Ghirardi, Rimini, \& Weber,
1986; Ghirardi, 2002)
attempt to resolve
this problem by modifying the unitary dynamics. In the GRW theory, 
there is a certain very small probability that the wave
function of a particle (the quantum state with respect to the
position basis in Hilbert space)
  will spontaneously collapse to a peaked
Gaussian of a specified width. For a macroscopic system consisting of many
particles, this probability can be close to 1 for very short time
intervals. In effect, GRW modify the unitary dynamics of standard 
quantum mechanics by adding uncontrollable noise. When the stochastic 
terms of the GRW dynamics become important at the mesoscopic and 
macroscopic levels, they tend to localize the wave function in space. 
So
measurement interactions involving macroscopic pieces
of equipment can be distinguished from elementary quantum processes,
insofar as they lead to the almost instantaneous collapse of the wave 
function and the correlation of the 
 measured observable with the position of a localized macroscopic pointer 
 observable.

The GRW dynamics for the density operator is a completely positive 
linear map. (See Simon, Buzek, \& Gisin, 2001, especially footnote 14. 
I am indebted to Hans Halvorson for bringing this point to my attention.) 
It follows that a GRW theory is empirically equivalent to a quantum 
theory with a unitary dynamics on a larger Hilbert space. Such a 
quantum theory will involve `hidden' ancillary degrees of freedom 
that are traced over. Since the GRW noise is uncontrollable in principle, 
there will be entangled states associated with this larger Hilbert space 
that cannot be prepared, and so cannot be exploited for steering in 
Schr\"{o}dinger's sense. This suggests that unconditionally
secure bit commitment would, in
principle, be possible via a protocol that requires Alice or Bob to access 
these hidden degrees of freedom in order to cheat. To put the point 
differently: unconditionally secure bit commitment is 
possible in the sort of quantum-like theory 
considered by Schr\"{o}dinger, because entangled pure states of a 
composite system 
collapse to proper mixtures as the component systems separate, which 
makes cheating via steering impossible. 
Similarly, in a GRW theory, the possibility of cheating via steering
is diminished to the extent that GRW noise cannot be controlled, and spontaneous 
collapse destroys or degrades nonlocal entanglement involving
inaccessible `hidden' degrees of freedom. So it seems that
 the GRW theory 
conflicts with the `no unconditionally secure bit commitment' 
information-theoretic 
constraint.

The other way of resolving the measurement problem---the `no-collapse'
route---is to keep the
linear dynamics and change the usual rule that associates a
specific catalogue of properties with a system via the quantum state
(the properties assigned unit probability by the state).\footnote{The
rule---often formulated for pure states as the `eigenvalue-eigenstate 
rule,' the assumption that an observable has a
definite value (so that the system has a definite property) if and only if
the state of the system is an eigenstate of the
observable---is explicit in von Neumann (1955, p. 253) and Dirac
(1958, pp. 46--47), and in
the EPR argument,
but notably
absent in Bohr's complementarity interpretation.}
This is tricky to do, because a variety of foundational
theorems severely restrict the assignment of properties or
values to observables under
  very general assumptions about the algebra of observables
(Kochen \& Specker, 1957), or restrict the assignment of
values to observables consistent with the quantum statistics (Bell,
1964).
The Bub-Clifton theorem (Bub \& Clifton, 1996) says that if
you assume that the family of definite-valued observables has a
certain structure (essentially allowing the quantum statistics to be
recovered in the usual way as measures over different possible definite
values or properties), and the pointer observable in a
measurement process belongs
to the set of definite-valued observables, then the class of such
theories---so-called `modal interpretations'---is uniquely specified.
This amounts to the requirement that the `no-collapse' theory should
include a mechanical account of the functioning of measuring
instruments. It turns out that such theories are characterized by a `preferred
observable' that always has a definite value. Different theories
involve different ways of selecting the preferred observable. For
example, the orthodox interpretation that leads to the measurement
problem can be regarded as a modal interpretation in which the
preferred observable is simply the identity $I$, and
Bohmian mechanics (Goldstein, 2001) can be regarded as a modal
interpretation in which the preferred observable is position in
configuration space.

In modal interpretations, measuring instruments generally do not function as
devices that faithfully measure dynamical quantities. In Bohmian 
mechanics, for example, what we call the
measurement of the $x$-spin of an electron which is in an eigenstate
of $z$-spin is not the measurement of a
property of the electron. Rather, an $x$-spin measurement involves a
certain dynamical evolution of the wave function of the electron in the
presence of a magnetic field, in which the wave function develops two
sharp peaks, one of which contains the electron. For a multi-particle
system, since the dynamical evolution depends on the position of the
system in
configuration space and the value of the wave function at that
point, the outcome of a spin measurement on one particle
will depend on the configuration of the other particles.

An alternative `no-collapse' solution to the measurement problem
is provided by the
many-worlds interpretation, first proposed by Everett (1957).
On this interpretation,
all the terms in a superposed or entangled quantum state (with
respect to a preferred basis) are regarded as actualized in different
worlds in a measurement, so
every possible outcome of a measurement occurs in some world. For
example, the measurement of the $x$-spin of an electron in an eigenstate of
$z$-spin is not a process that reveals a pre-existing spin value;
rather, it is a process in which an indefinite spin value becomes
definite with different spin values in different worlds. There are 
a variety of different versions of Everett's interpretation in the 
literature (see Wallace, 2003 for a recent discussion). On Bell's
characterization (Bell, 1987b), the many worlds interpretation
is presented as equivalent to Bohmian mechanics without 
the particle trajectories.

A modal interpretation or `no collapse'
hidden variable theory is proposed as a (`deeper')
mechanical theory underlying the statistics of a
$C^{*}$-algebraic quantum theory or its Hilbert space representation that
includes
a mechanical account of our measuring
instruments as well as the phenomena they reveal, i.e., as an 
extension of a quantum theory. From the CBH
theorem, a theory satisfies the information-theoretic constraints if 
and only if it is empirically equivalent to a quantum theory (a 
theory where the observables, the states, and the dynamics are 
represented as outlined at the end of Section 2). 
So, given the
information-theoretic constraints, any empirically adequate extension 
of a quantum theory in this sense must be empirically equivalent to the 
quantum theory. 

Consider Bohmian mechanics as an example. The additional mechanical
structures postulated as underlying the quantum statistics in Bohmian 
mechanics 
are the particle trajectories in
configuration space, and the wave function as a guiding field (which 
evolves via the Schr\"{o}dinger equation). 
The dynamical evolution of a
Bohmian particle is described by a deterministic
equation of motion in configuration space that is guaranteed to produce the
quantum statistics for all quantum measurements,
if the initial distribution over particle positions (the hidden
variables)
is the Born distribution. The Bohmian algebra of 
observables is the commutative algebra generated by the position observable 
and the Bohmian particle dynamics is nonlinear, so Bohmian mechanics is not a 
quantum theory in the sense of the CBH theorem.
In Bohmian mechanics, the Born distribution is treated as an
equilibrium distribution, and non-equilibrium distributions can be
shown to
yield predictions that conflict with the information-theoretic
constraints.
Valentini (2002) shows how nonequilibrium distributions can be
associated with
such phenomena as instantaneous signalling between spatially
separated systems and the
possibility of distinguishing nonorthogonal pure states
(hence the possibility of cloning such states). Key distribution
protocols whose security depends on `no information gain without
disturbance' and `no cloning' would then be insecure against attacks
based on exploiting such non-equilibrium distributions. So, in 
Bohmian mechanics,
the fact that the information-theoretic constraints
hold depends on (and, in this sense, is explained by) 
a contingent feature of the theory:
that the universe
has in fact reached the equilibrium
state with respect to the distribution of hidden variables.

But now it is clear that there can be no empirical evidence for the
additional mechanical elements of Bohmian mechanics that would not 
also be evidence for the statistical predictions of a quantum theory, 
because such evidence
is unobtainable in the equilibrium state. If the
information-theoretic constraints apply at the phenomenal
level then, according to Bohmian mechanics, the universe must be in
the equilibrium state, and in that case there can be no evidence for
Bohmian mechanics that goes beyond the empirical content of a quantum
theory (i.e., the statistics of quantum superpositions and entangled states).
Since it follows from the CBH theorem that a similar analysis will apply to
any `no collapse' hidden variable theory or modal interpretation,
there can, in principle, be no empirical grounds for choosing among these
theories, or between any one of these theories and a
quantum theory.

\section{The Completeness of Quantum Theory}

What is the rational epistemological stance in this situation?
Consider the case of thermodynamics, which is a theory formulated in terms
of constraints at the phenomenal level (`no perpetual motion machines of
the first and second kind'), and the kinetic-molecular theory, which is a
statistical mechanical theory of processes at the microlevel
that provides a mechanical
explanation of why thermodynamic phenomena are constrained by the
principles of thermodynamics.
Should we take the ontology of the kinetic-molecular theory
seriously as a realist explanation of observable thermodynamic phenomena?
This
was regarded as an open question at the turn of the 20th century 
before Perrin's
(1909) experiments on Brownian motion.
(For an account see Nye (1972).) Why? Because before these experiments
there was no empirical \textit{scale} constraint on the sizes of molecules or
atoms, the basic
structural elements of the kinetic-molecular theory. So there were no
empirical
grounds for taking the unobservable aspects of the ontology proposed
by the kinetic theory seriously as an explanation of the observable
phenomena.
To put it simply: you ought to be able to count the number of molecules
on the head of a
pin, or you might as well be talking about angels.

It was Einstein's analysis of Brownian motion and his prediction of
observable fluctuation phenomena that allowed the crucial
scale parameter, Avogadro's number, to be pinned down.
Without the possibility of observable fluctuation phenomena, the
kinetic theory would have been, to use Poincar\'{e}'s phrase, no more than
a `useful fiction' (Poincar\'{e}, p. 1912):
\begin{quote}
\ldots\ the long-standing mechanistic and atomistic hypotheses have
recently taken on enough consistency to cease almost appearing to us
as hypotheses; atoms are no longer a useful fiction; things seem to us
in favour of saying that we see them since we know how to count
them. \dots\ The brilliant determination of the number of atoms made
by M. Perrin has completed this triumph of atomism. \ldots\ The atom
of the chemist is now a reality.
\end{quote}

Einstein's first paper on Brownian motion
(Einstein, 1956, pp. 1--2) makes
a similar point\footnote{I have used Penelope Maddy's translation for the
last sentence Maddy. 1997, p. 139). The English version has `had' for `should'
and `proved'
for `prove.' The German reads: `Erwiese sich umgekehrt die Voraussage
dieser Bewegung als unzutreffend, so w\"{a}re damit ein schwerwiegendes
Argument gegen die molekularkinetische Auffassung der W\"{a}rme gegeben.'
See Nye (1997, p. 139)}:
\begin{quote}
In this paper it will be shown that
according to the molecular-kinetic theory of heat, bodies of
macroscopically-visible size suspended in a liquid will perform
movements of such magnitude that they can be easily observed in a
microscope, on account of the molecular motions of heat. \ldots\
If the movement discussed here can actually be
observed (together with the laws relating to it that one would expect
to find), then classical thermodynamics can no longer be looked upon as
applicable with precision to bodies even of dimensions distinguishable
in a microscope: an exact determination of actual atomic dimensions is
then possible. On the other hand, should the prediction of this
movement prove to be incorrect, a weighty argument would be provided
against the molecular-kinetic theory of heat.
\end{quote}

Compare, now, the kinetic-molecular
theory relative to thermodynamics, and
a modal interpretation or `no collapse' hidden variable
theory, which is proposed as an extension of a quantum theory to 
solve the measurement problem and
provide an answer to the question: How could the world possibly be the 
way a quantum theory says it is? From the CBH theorem, this is 
amounts to asking: How is it possible that the information-theoretic 
constraints hold in our world? To focus the discussion,
consider Bohmian mechanics.
The additional mechanical elements of
Bohmian mechanics are the Bohmian particle
trajectories in configuration space and the wave function as guiding 
field (the quantum
state in configuration space). In Bohmian mechanics,  a
measurement is represented by a dynamical evolution induced by a 
measurement interaction in the configuration space of the combined 
system plus measuring instrument. A Stern-Gerlach measurement of 
the $x$-spin of a
spin-$\frac{1}{2}$ particle in an eigenstate of $z$-spin is a 
particularly simple example, since here the position of the particle 
functions as the measurement `pointer' for the spin value. The 
measurement is represented by the dynamical
evolution of the
particle in configuration space (which, in this special case, is
just real space) under the influence of a guiding field
represented by the wave function evolving in the presence of an 
inhomogeneous magnetic field.
During the measurement process, the wave function evolves
  in such a way as to entangle the
position of the particle---in effect, the measurement `pointer'---and the spin.
That is, the wave function develops two peaks correlated with the
two possible spin eigenstates. Since, by assumption, the Bohmian particle
always has a definite
position, which must be in one or the other of the two peaks,
this position value
(measured as either `up' or `down' in the case of a Stern-Gerlach
measurement of $x$-spin) can be associated with a particular spin
eigenvalue. The remaining term in the entangled state can be dropped,
because it plays a negligible role in determining the future motion
of the particle. So there is an `effective collapse' of the wave
function (see Maudlin, 1995).

It follows from the Bohmian particle dynamics and the Schr\"{o}dinger 
evolution for the guiding field that the distribution of particle
positions after any measurement (as given by the effective wave 
function) will never vary from
the equilibrium Born distribution if the initial distribution is the Born
distribution, so there can be no observable `fluctuation
phenomena' analogous to the observable fluctuation phenomena of
Brownian motion in the thermodynamics case. This means
that there can be no empirical constraint on the Bohmian particle
trajectories analogous
to the empirical scale constraint in the case of the kinetic-molecular theory
(if
our universe is indeed in the equilibrium state, when the
information-theoretic constraints apply).

If it was correct to suspend judgement about the
reality of atoms before Perrin's experiments, the correct conclusion
to draw with respect to Bohmian mechanics is that, since---\textit{in principle}---
there can be no empirical grounds for taking the unobservable
Bohmian trajectories seriously as an explanation of observable
quantum phenomena (assuming our universe is in the equilibrium
state), Bohmian mechanics
  is, at best, a
`useful fiction' in Poincar\'{e}'s sense. (`Useful' here only in
satisfying a philosophical demand for the sort of explanatory completeness
associated with commutative theories, in that the theory provides a
mechanical account of quantum phenomena, including an account of the
measuring instruments that reveal these phenomena.)

Note that the argument here is not that it is never rational to believe a 
theory over an empirically equivalent rival: the methodological 
principle I am appealing to is weaker than this.  Rather, my point is 
that if $T'$ and $T''$ are empirically equivalent extensions of a 
theory $T$, and if $T$ entails that, in principle, there \textit{could 
not be} evidence 
favoring one 
of the rival extensions $T'$ or $T''$, 
then it is not rational to believe either 
$T'$ or $T''$.  

To clarify this point (following a suggestion by Hans Halvorson): 
Say that $T$ and $T'$ are \textit{weakly empirically equivalent} in a world $W$ 
just in case 
the theories are equivalent relative to all evidence available in $W$. 
And say that $T$ and $T'$ are \textit{strongly empirically equivalent} 
just in case 
they are weakly empirically equivalent in all possible worlds 
(in other words, there could not possibly be evidence favoring one 
theory over its rival), where the set of possible worlds is 
determined by an accepted physical theory. Now let $T$ be a quantum theory, and 
let $T', T'', \ldots$ 
be various extensions of this quantum theory (e.g., Bohm, Everett, etc.). 
If we accept $T$, then (by the CBH theorem) 
we accept that there could be no evidence favoring 
any one of the theories $T', T''$ as a matter of physical law.  
In other words, we accept that 
there is no possible world satisfying the information-theoretic 
constraints in which there is evidence favoring one of 
these extensions over its rivals.

Now, strictly speaking, thermodynamics
is falsified by the kinetic-molecular theory: matter is `grainy,' and
the second law has only a statistical validity. The phenomena that 
reveal the graininess in the
thermodynamics case are fluctuation
phenomena, and these are (small) departures from equilibrium. So, one 
might argue, the
appropriate case to consider for a quantum theory is not the
equilibrium version of Bohm's
theory, but rather
the non-equilibrium version.

I grant that it could turn out to be false that the
information-theoretic constraints hold in our universe and that some 
day we will find experimental evidence that conflicts with the 
predictions of a quantum theory (in which case 
the nonequilibrium version of Bohm's theory might turn out to be
true). The relevant point about the thermodynamics case is that the
kinetic-molecular theory was regarded as only a `useful fiction'
before Einstein showed that the theory could have excess empirical content over
thermodynamics (even though acceptance of the theory
ultimately required a revision of the principles of thermodynamics). 
The methodological
moral I draw from the thermodynamics case is
simply that a mechanical theory that purports to solve the measurement
problem is not acceptable if it can be shown that, \textit{in principle}, the
theory can have no excess
empirical content over a quantum theory. By the CBH theorem, given the 
information-theoretic constraints any 
extension of a quantum theory, like Bohmian mechanics, must be 
empirically equivalent to a quantum theory, so no such theory 
can be acceptable as a deeper mechanical explanation of why 
quantum phenomena are subject to the information-theoretic constraints. 
To be acceptable, a mechanical theory that includes an account
of our measuring instruments as well as the quantum phenomena they reveal
(and so purports to solve the measurement problem) \textit{must violate one or
more of the information-theoretic constraints.}

Similar remarks apply to other `no collapse' hidden variable 
theories or modal interpretations,
including the many worlds interpretation: by the CBH theorem, the
additional mechanical elements of these theories must be idle if the
information-theoretic constraints apply.
I conclude that the rational
epistemological stance is to suspend judgement about all these
empirically equivalent but necessarily underdetermined theories and 
regard them all as unacceptable.
\textit{It
follows that
our measuring instruments ultimately remain black boxes at some
level} that we represent in the theory simply as probabilistic 
sources of ranges of labelled events or `outcomes,' i.e., effectively 
as sources of signals, where each signal is produced with a certain 
probability. But this amounts to treating
a quantum theory as \textit{a theory about the
representation and manipulation of information} constrained by
the possibilities and
impossibilities of information-transfer in our world (a fundamental
change in the aim of physics), rather than a theory
about the ways in which nonclassical waves or particles move.
The explanation for the impossibility of a `detached observer'
description then lies in the constraints on
the representation and manipulation of information that
hold in our world.

So a consequence of rejecting
Bohm-type hidden variable theories or other `no collapse' theories
 is that we recognize 
information as a new sort of
physical entity, not reducible to the motion of particles or fields.
An entangled
state should be thought of as a nonclassical communication channel
that we have discovered to exist in our quantum universe, i.e., 
as a new sort of nonclassical `wire.' We can use
these communication channels to do things that would be impossible
otherwise, e.g., to teleport states, to compute in new ways, etc. A quantum
theory is then about the properties of these communication channels, and
about the representation and manipulation of states as sources of information
in this physical sense.

Just as the rejection of Lorentz's theory in favour of special relativity
(formulated in terms of Einstein's two principles: the equivalence of
inertial frames for all physical laws, electromagnetic as well as
mechanical, and the constancy of the velocity of light in vacuo for
all inertial frames) involved taking
the notion of a field as a new physical primitive, so the rejection of
Bohm-type hidden variable theories in favour of quantum 
mechanics---characterized via the CBH
theorem in terms of three information-theoretic principles---involves
taking the notion of quantum information as a
new physical primitive. That is, just as Einstein's analysis
(based on the assumption that we live
in a world in which natural processes are subject to certain
constraints specified by the principles of special relativity)
shows that we do not need the mechanical structures in Lorentz's
theory (the aether, and the behaviour of electrons in the aether)
to explain electromagnetic phenomena, so the CBH analysis
(based on the assumption that we live in a world in which there are
certain constraints on the acquisition, representation, and
communication of information) shows that we do not need the
mechanical structures in Bohm's theory
(the guiding field, the behaviour of particles in the guiding field)
to explain quantum phenomena. You can, if you like, tell a story along
Bohmian, or similar, lines (as in other `no collapse' interpretations)
but, given the information-theoretic constraints, such a story can, 
in principle,
have no excess empirical content over
Hilbert space quantum mechanics (just as Lorentz's theory, insofar as it is
constrained by the requirement to reproduce the empirical content of the
principles of special relativity, can, in principle, have no excess
empirical content over Einstein's theory).

Something like this view
seems to be implicit in Bohr's complementarity interpretation of
quantum theory. For
Bohr, quantum mechanics is complete and there is no measurement
problem, but measuring instruments ultimately remain outside the
quantum description: the placement of the `cut' between system and measuring
instrument is arbitrary, but the cut must be placed somewhere.
Similarly, the argument here is that, if the information-theoretic
constraints hold in our world, the measurement problem is a
pseudo-problem, and the whole idea of an empirically equivalent
`interpretation' of quantum
theory that `solves the measurement problem' is to miss the point of
the quantum revolution.

From this
information-theoretic perspective, the relevant `measurement problem' 
is how to account for the emergence of classical information, 
the loss of interference and 
entanglement, when we perform quantum measurements. The solution to this problem 
appears to lie in the phenomenon of environmental decoherence 
that occurs during a quantum measurement. In effect, we design 
measurement instruments to exploit decoherence: an 
instrument-environment interaction that results almost 
instantaneously (as a result of information loss to the environment) 
in a particular sort of noisy entanglement between the measured 
system, the measuring instrument, and the environment. The noisy 
channel means that the system, monitored by the measuring 
instrument, behaves classically: all the subsequent information-processing 
we can do with it will be classical. Technically, 
the von Neumann entropy
measuring quantum information reduces to the classical Shannon entropy under
the loss of information induced by
decoherence. So most of the 
information in a quantum state that can be processed is not accessible 
in a measurement---just one bit of the potentially infinite amount of
  quantum information in a spin-$\frac{1}{2}$ system, for example, can be
accessed in a measurement of spin in a particular direction:
the classical information content of
the two alternative spin values associated with that direction. 

The standard measurement problem is the problem of showing that
after a measurement interaction the measured system is actually in one of the
eigenstates of the measured observable, with the appropriate quantum
mechanical probability (which reflects our ignorance of the actual
eigenstate before the measurement), and that the measured observable therefore
has a definite value (according to the usual interpretation that
takes the definite or determinate properties of a system as the
properties assigned unit or zero probability by the state). That is, 
the standard measurement problem is the problem of accounting for the
definiteness or determinateness of pointer readings and measured 
values in a measurement process. John Bell (1990) 
famously objected to appealing to decoherence as a `for all practical 
purposes (FAPP)' solution to this problem. What he objected to was the 
legitimacy of regarding the pointer 
observable and the measured observable correlated with the pointer
 as having definite values, on the basis that decoherence justifies 
tracing over the environment  and ignoring 
certain correlational information in the 
 system-instrument-environment entangled state, for all practical purposes. 
 Bell rightly objected that decoherence cannot
guarantee the determinateness of properties in this way, and
that a FAPP solution to the problem cannot therefore underwrite a
quantum ontology for a fundamental `detached observer' 
mechanical theory of events
at the microlevel. But the objection does not apply to the problem of
accounting for the emergence of classical information in quantum
measurements.

`Why the quantum?' was one of John Wheeler's
`Really Big Questions.' The characterization of quantum mechanics in
terms of three information-theoretic constraints provides an answer to this
question: a quantum theory is fundamentally a theory about the
possibilities and impossibilities of information transfer in our
world, given certain constraints on the acquisition,
representation, and communication of information,
not a theory about the mechanics of nonclassical waves or particles.
In the debate between Bohr and Einstein on the interpretation of
quantum theory, this answer to Wheeler's question sides with Bohr.

The focus on quantum information as an answer to Wheeler's question
about the quantum  has been impressively successful in
terms of new physics
over the past twenty years or so.
Where Einstein and Schr\"{o}dinger saw a problem
(e.g., the nonlocality of entanglement in the EPR experiment), 
contemporary physicists
see an opportunity to exploit entanglement as a new sort of 
nonclassical communication channel
(e.g., for teleportation, or for new modes of communication and computation).
This is a major revolution in the aim and practice of physics. As
Andrew Steane (1998) puts it:
\begin{quote}
Historically, much of fundamental physics has been concerned with
discovering the fundamental particles of nature and the equations
which describe their motions and interactions. It now appears that a
different programme may be equally important: to discover the ways
that nature allows, and prevents, information to be expressed and
manipulated, rather than particles to move.
\end{quote}

\section*{Acknowledgements}

Comments on a first
  draft of this paper by Guido Bacciagaluppi, Jossi Berkovitz, Michael Cifone, 
  William Demopoulos, Hans Halvorson,
  Michael Silberstein, Jos Uffink, and an anonymous referee were very helpful
  in shaping the reformulation of the arguments in this version. The
  ideas in this paper originated during a research leave 
  supported
  by a University of Maryland Sabbatical Leave Fellowship and a
  General Research Board Fellowship in 2001--2002.

\end{document}